\documentstyle[epsfig]{mn}
\begin{document}
\LARGE
\normalsize

\title[GRS~1915+105]
{Giant repeated ejections from GRS 1915+105}
\author[R.~P.~Fender \& G.~G.~Pooley ]
{R. P. Fender$^1$ and 
G. G. Pooley$^2$  \\
$^1$ Astronomical Institute `Anton Pannekoek', University of Amsterdam,
and Center for High Energy Astrophysics, Kruislaan 403, \\
1098 SJ, Amsterdam, The Netherlands {\bf rpf@astro.uva.nl}\\
$^2$ Mullard Radio Astronomy Observatory, Cavendish Laboratory,
Madingley Road, Cambridge CB3 OHE {\bf ggp1@cam.ac.uk}\\
}

\maketitle

\begin{abstract}

We report simultaneous millimetre and infrared observations of a
sequence of very large amplitude quasi-periodic oscillations from the
black hole X-ray binary GRS 1915+105. These oscillations are near the
end of a sequence of over 700 repeated events as observed at 15 GHz,
and are
simultaneous at the mm and infrared
wavelengths to within our time resolution
($\leq 4$ min), consistent with the respective emitting regions being
physically close near the base of the outflow.  One infrared event
appears to have no mm counterpart, perhaps due to highly variable
absorption. The overall radio--mm--infrared spectrum around the time
of the observations does suggest some absorption at lower
frequencies. We calculate the energy and mass-flow into the outflow
for a number of different assumptions, and find that the time-averaged
power required to produce the observed synchrotron emission cannot be
much less than $3 \times 10^{38}$ erg s$^{-1}$, and is likely to be
much larger.  This minimum power requirement is found regardless of whether
the observed emission arises in discrete ejections or in an internal
shock in a quasi-continuous flow. Depending on the similarity of the
physical conditions in the two types of ejection,
GRS 1915+105 may be supplying more power (and mass,
if both have the same baryonic component) to the jet during periods of
repeated oscillations than during the more obvious larger events.

\end{abstract}


\begin{figure*}
\centering
\leavevmode\epsfig{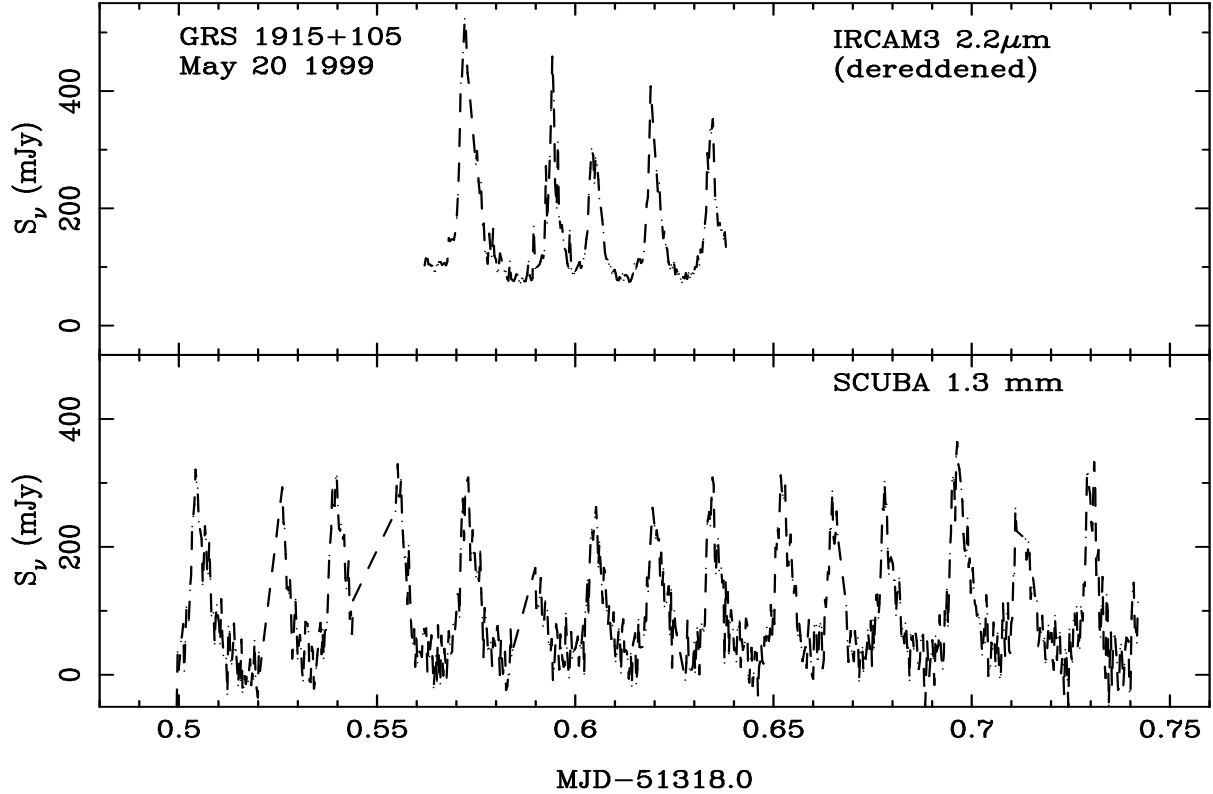}
\caption{Simultaneous mm (1.3~cm) and infrared K-band
(2.2~$\mu$m) light curves of GRS~1915+105. The infrared data
have been dereddened by $A_{\rm K} = 3.3$~mag. These are the largest
repeated oscillations ever observed from GRS 1915+105.}
\label{}
\end{figure*}

\section{Introduction}

The `microquasar' GRS 1915+105 is one of the most celebrated and
widely-studied astrophysical objects of recent years. The system is
extremely luminous and variable in both hard and soft X-rays
(e.g. Foster et al. 1996; Morgan, Remillard \& Greiner 1997; Belloni
et al. 2000) and is a source of relativistic jets observed on arcsec
and milliarcsec angular scales (Mirabel \& Rodr\'\i guez 1994,
hereafter MR94; Fender et al. 1999, hereafter F99; Rodr\'\i guez \&
Mirabel 1999, hereafter RM99; Dhawan, Mirabel \& Rodr\'\i guez 2000).
Sams, Eckart \& Sunyaev (1996) have reported extended infrared
emission from GRS 1915+105, but its relation to the radio ejections is
at present unclear.

X-ray dips on timescales of minutes have been interpreted by Belloni
et al. (1997a,b) as the repeated disappearance and refill of the inner
accretion disc, possibly due to extremely rapid transitions between
`canonical' black hole accretion states (Belloni et al. 2000). Pooley
\& Fender (1997; hereafter PF97) reported radio oscillations
associated with such dips, and Fender et al. (1997; hereafter F97)
discovered infrared analogs of these oscillations. The flat spectrum
and correlated radio : infrared behaviour suggested that nonthermal
synchrotron emission extended from the radio to the infrared regimes,
the first time such high-frequency synchrotron emission had been
observed from an X-ray binary (F97). Combined with the unstable
accretion disc model of Belloni et al. (1997a,b) we suggested that a
fraction of the inner disc was being repeatedly accelerated and
ejected from the system (F97; PF97). Eikenberry et al. (1998a)
confirmed the association between X-ray and infrared events, and
Mirabel et al. (1998, hereafter M98) clearly observed the correlation
between X-ray, infrared and radio behaviour in the source. Additional
simultaneous observations (Fender \& Pooley 1998, hereafter FP98)
showed a very clear correlation between sequences of oscillations at
radio and infrared wavelengths. Delays between different radio bands
(PF97; M98) and between the infrared and radio bands (M98; FP98)
clearly indicate that optical depth effects play an important role in
the observed emission from these ejections.

Eikenberry et al. (1998b) showed that infrared emission
line strengths vary in proportion to the continuum during
oscillations. A single X-ray dip, spectrally associated with the
temporary disappearance of the inner accretion disc, was also found to
coincide with a small radio flare (Feroci et al. 1999).  More recently
Eikenberry et al. (2000) report faint infrared flares whose
association with the X-ray behaviour is uncertain, and Ogley et al.
(2000) report significant flux from GRS 1915+105 at sub-millimetre
wavelengths. 

\section{Observations}

GRS 1915+105 was observed simultaneously on 1999 May 20 with the
United Kingdom Infrared Telescope (UKIRT) and the James Clerk Maxwell
Telescope (JCMT), both located on Mauna Kea, Hawaii.

\subsection{UKIRT}

GRS 1915+105 was observed with IRCAM3 in the infrared K-band ($2.2
\mu$m) on 1999 May 20, simultaneously with
the longer duration of JCMT SCUBA observations (see below). Data
reduction and calibration were performed with {\sc iraf}, along
the lines described in F97. Five clear oscillation events were
detected.  The {\em undereddened} infrared flux densities reached 25
mJy at the peak of the oscillations, the largest amplitude
oscillations reported to date in the infrared. The data are plotted in
Figs 1 \& 2, dereddened by $A_{\rm K} = 3.3$ mag (F97;
this value is still rather uncertain).

\begin{figure}
\leavevmode{\epsfig{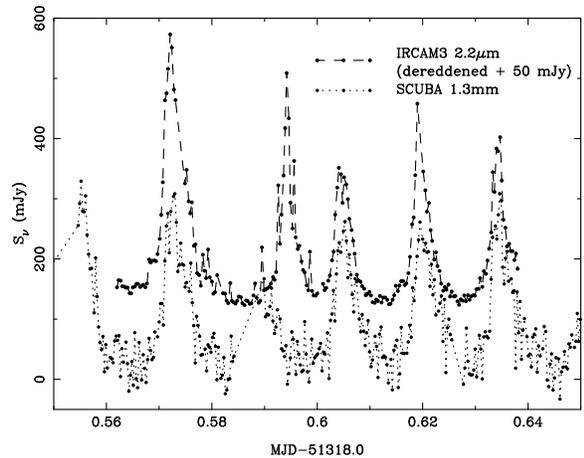}}
\caption{
Enlargement of the simultaneous mm/infrared observations. While four
of the five infrared oscillations correlate well with the mm emission,
one event is clearly anomalous, with little mm response to the second
infrared oscillation.
}
\end{figure}

\subsection{JCMT}

The 1350 $\mu$m detector of the SCUBA instrument (Holland et al. 1999)
on JCMT was used in the photometry. Each integration lasted
approximately 4 min. Calibration of the flux-density scale used
observations of Mars and Uranus.  The airmass ranged from 1.01 to 1.92
during the observations, and the optical depth at 1350 $\mu$m was less
than 0.2 throughout. The data are plotted in Figs 1 \& 2.

\subsection{Radio}

In order to piece together the composite radio--mm--infrared spectrum
of the source at the epoch of our observations, we have utilised radio
data from two different monitoring programs. Firstly we have used
public data at 2.3 \& 8.3 GHz from the Green Bank Interferometer (GBI)
monitoring program (e.g. Waltman et al. 1994).  Observations at 15
GHz with the Ryle Telscope (RT, e.g. PF97) reveal
strong oscillations for at least 9 days before and 2 days after the
simultaneous UKIRT/JCMT observations (Fig 3), with a slowly rising
trend in mean level and amplitude.


\begin{figure*}
\leavevmode{\epsfig{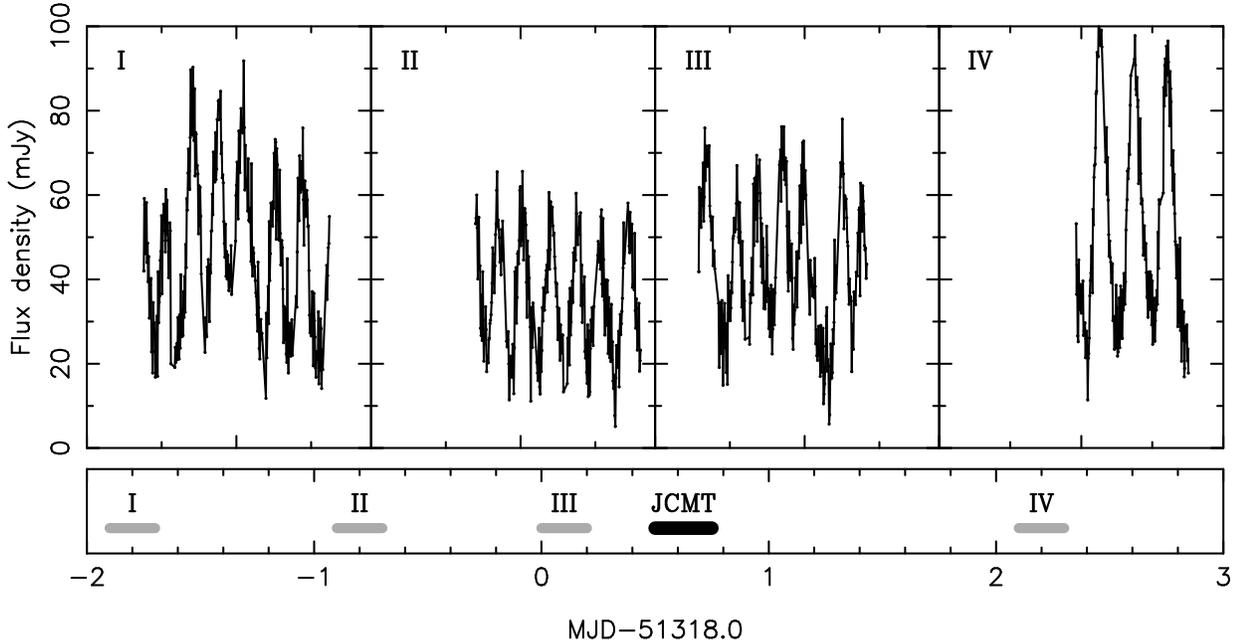}}
\caption{Ryle Telescope observations of GRS 1915+105 before and after
the JCMT/UKIRT observations. Each of the top panels corresponds to 
0.2 days, and their distribution with respect to the
JCMT/UKIRT observations is indicated in the lower panel. It is clear
that strong radio oscillations were occurring before and after the
JCMT/UKIRT observations, with a similar quasi-period, but lower
amplitude. In fact the RT observations show that the oscillations were
probably continuous between MJD 51309--51320.
}
\end{figure*}

\section{Temporal and spectral behaviour}

The JCMT observations reveal a sequence of 15 
millimetre-wavelength oscillations with a quasi-period of $\sim 23$
min. Four of the oscillations have been observed simultaneously in the
near-infrared, and both the (dereddened) near infrared and mm
oscillations have an amplitude of 300--350 mJy. These are by far the
largest amplitude oscillations ever observed from GRS 1915+105,
including radio wavelengths (see e.g. PF97 for
`typical' radio oscillations). There is no detectable delay between
the emission at the two wavelengths, to the time resolution of the
JCMT data, $\sim 4$ min.

F97, M98 and FP98
all report infrared oscillations of comparable amplitude 
(when dereddened by $A_{\rm K} = 3.3$ mag) to radio oscillations
observed around the same time. Although radio observations were not
made strictly simultaneously with these mm/infrared observations,
monitoring with the RT clearly reveals radio oscillations, with a
comparable quasi-period of $\sim 20$ min, to have been occurring for
at least two days before and afterwards (Fig 3). However, the
amplitude of the mm and infrared oscillations is 
about {\em five times} greater than that observed at radio
wavelengths. The mean radio -- infrared spectrum for the four-day
period illustrated in Fig 3 is shown in Fig 4.

A striking feature of the light curves in Figs 1 \& 2
is the infrared oscillation which starts around MJD 51318.59 which
does not appear to have a mm counterpart, unlike the other four
simultaneously observed events. We have carefully checked the data
reduction techniques to see if this was due to human error, but found
no evidence of this. We 
have no clear physical intepretation of this
phenomena, except to suggest that it was due to strong and variable
absorption which only signficantly affected the lower-frequency
emission (for example the optical depth to free-free absorption, $\tau
\propto \nu^{-2.1}$).  As noted above, the overall radio--mm--infrared
spectrum at the time of these observations (Fig 4) was steeper than
previously observed, perhaps also indicative of some absorption. If
this is a correct interpretation of the `failed' mm event, then we
would have expected the radio emission to have been completely
absorbed at this time also.

\section{Energetics and mass outflow rate}

\begin{table*}
\begin{tabular}{llllll|ll}
\hline
Case & f & L(erg) & B$_{\rm eq}$(G) & $E_{\rm min}$(erg) & M (g) & P (erg s$^{-1})$ & $\dot{M}_{\rm jet}$(g s$^{-1}$) \\ 
\hline
e$^+$:e$^-$, $\Gamma=1$ & 0.01
& $3 \times 10^{37}$ & 145 & $6 \times 10^{40}$ & -- & $5 \times
10^{37}$ & -- \\
e$^+$:e$^-$, $\Gamma=1$ & 0.1
& $3 \times 10^{37}$ & 75 & $2 \times 10^{41}$ & -- & $2 \times
10^{38}$ & -- \\
e$^+$:e$^-$, $\Gamma=1$ & 1.0
& $3 \times 10^{37}$ & 40 & $4 \times 10^{41}$ & -- & $3 \times
10^{38}$ & -- \\
e$^+$:e$^-$, $\Gamma=5$ & 1.0
& $4 \times 10^{39}$ & 115& $3 \times 10^{43}$ & -- &
$3 \times 10^{40}$ & -- \\
p$^+$:e$^{-}$, $\Gamma=1$ & 1.0
& $3 \times 10^{37}$ & 40 & $4 \times 10^{41}$ & $2 \times 10^{23}$ &
$3 \times 10^{38}$ & $2 \times 10^{20}$ \\
p$^+$:e$^{-}$, $\Gamma=5$ & 1.0
& $5 \times 10^{39}$ & 115& $1 \times 10^{46}$ & $3 \times 10^{24}$ &
$8 \times 10^{42}$ & $4 \times 10^{21}$ \\
\hline
\end{tabular}
\caption{Calculation of radiative luminosity, equipartition magnetic
field, total energy and jet power \& mass-flow rate for the oscillations
reported here, given different physical assumptions. $\Gamma$ is the
bulk motion Lorentz factor, $f$ is the `filling factor'.
In these
calculations a distance of 11 kpc and
Doppler factors for relativistic bulk motion which are the same as
those reported in F99 are all assumed. Mass flow rate 
$\dot{M}_{\rm jet}$ and jet power P are
based upon one ejection every 20 min. For more details, see main text.}
\end{table*}

The radiative luminosity of these oscillations is large -- for a flat
spectrum of amplitude 300 mJy from 1 GHz to $1.4 \times 10^5$ GHz
($\equiv 2.2\mu$m), at a distance of 11 kpc, it is $3 \times 10^{37}$
erg s$^{-1}$ (the time-averaged radiative luminosity is around half
this value). As is the case for all synchrotron emitting plasmas for
which adiabatic expansion losses dominate, this is likely to be a
significant underestimate of the power being supplied to the jet.
Furthermore we assume the emission arises in a partially self-absorbed
jet which retains the same power-law distribution of electrons
(i.e. $p=2.6$ where $N(E)dE \propto E^{-p}dE$) as observed in
optically thin ejections (F99). In this situation the flat spectrum is
produced by a conical, partially self-absorbed jet
(e.g. Blandford \& K\"onigl 1979; Reynolds 1982). 

In addition several factors which can further affect the energy budget
are uncertain, in particular whether or not the small ejections share
the same bulk relativistic motions as the larger ejections, whether
each radiating electron has an associated cold proton, and what the
filling factor (ie. the effective volume) of the ejecta is.
The procedure for calculating the energy and mass of the ejections is
as follows:

\begin{itemize}
\item{Transform observed flux densities and frequencies back to their
rest frame (identical if no bulk relativistic motion).}
\item{Integrate rest-frame luminosity.}
\item{Calculate maximum emitting volume -- in this case based on the
five-minute rise time this is $3 \times 10^{39}$ cm$^3$. The effective
emitting volume is this volume multiplied by a `filling factor', $f$.}
\item{From the volume, spectrum and luminosity, calculate
equipartition magnetic field, and corresponding
minimum internal energy.}
\item{For baryonic case, add one proton for each electron.}
\item{For cases with bulk relativistic motion, add in kinetic energy
and multiply by two, under the assumption that observed emission was
dominated by one (approaching) component only.}
\item{Divide by repetition quasi-period of oscillations to obtain
time averaged energy and mass outflow rate.}
\end{itemize}

We have tabulated results for different cases in table 1; 
for bulk relativistic motion we have used the Doppler factors
corresponding to $\beta=0.98$, $\theta=66^{\circ}$ from F99. 

A significant constraint is that the lack of evidence for synchrotron
losses (based on the similarity of decay rates at widely different
wavelengths) at 2.2$\mu$m on a timescale of $\sim 10$ min, implies
that $B_{\rm max} \la 30$ G. For bulk motions with Doppler factor
$\delta$, this limit is shifted slightly to $B_{\rm max}
\delta^{-1/3}$ which, for $\delta=0.34$ in this case means the
limiting field is $\sim 40$G, instead of $\sim 30$G, not a major
difference.  Thus while reducing the emitting volume via the filling
factor (see Table 1) decreases the minimum energy, the stronger
derived equipartition magnetic field is irreconcilable with the
observed minimum lifetimes.  As already noted in F97, a field of order
10 G will cause a cut-off in the spectrum of the oscillations around
the optical band.  Reducing the magnetic field below the equipartition
value results in the internal energy being dominated by the electrons,
for which total energy $E \propto B^{-3/2}$. Because of this
constraint, the realistic minimum energy cannot be reduced much below
the value for a non-relativistic non-baryonic ejection with filling
factor $f=1.0$, which is $4 \times 10^{41}$ erg, with a corresponding
time-averaged power requirement of $3 \times 10^{38}$ erg
s$^{-1}$. This corresponds to a radiative efficiency for the outflow
of $\leq 5$\%.  
This minimum power requirement is not strongly 
affected by our assumption of the spectral form of the electron
distribution (e.g. for $p=2.0$ the minimum power is reduced by only a
factor of three).
Only in the case of baryonic ejections at high velocities
does the large number of lower energy electrons become
significant, as each has an associated proton. 
As a result it does not matter whether the emission
arises in a discrete `plasmon' or an internal shock in a quasi-steady
flow. It is interesting to compare the power for baryonic ejections
with $\Gamma=5$, $2 \times 10^{43}$ erg s$^{-1}$,
with that calculated for the same criteria for the `major' ejections
in F99, $2 \times 10^{39}$ erg s$^{-1}$. This is due to the observed
optically thin cm spectrum, $S_{\nu} \propto \nu^{-0.8}$, being
assumed in F99 to have no high-frequency excess, and therefore a much
lower integrated luminosity than the flat spectrum oscillations
reported here. Further prompt mm and infrared observations during
`major' outbursts are required to investigate this.

\begin{figure}
\leavevmode{\epsfig{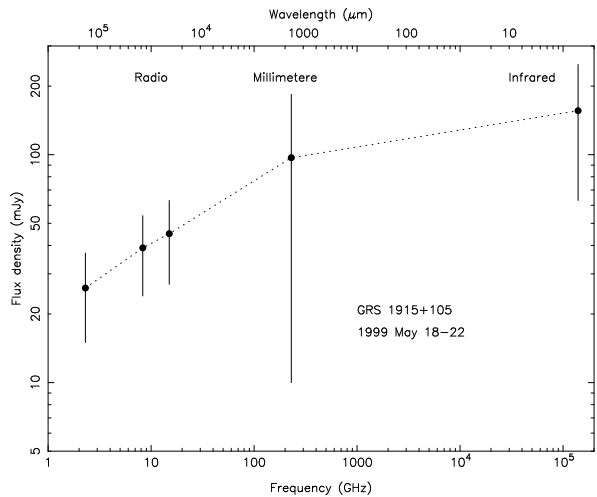}}
\caption{
The mean radio--mm--infrared spectrum of GRS 1915+105 in the interval
MJD 51316--51320, centred on our simultaneous infrared/mm
observations. The spectrum appears to be significantly inverted at
longer wavelengths. Vertical bars on each point are the statistical
standard deviation for the data set, and reflect the relative
amplitude of observed variability at each wavelength ({\em not}
measurement errors).
}
\end{figure}

\section{Discussion}

M98 have shown that the wavelength-dependent time
delays (radio--radio and radio--infrared) observed from GRS 1915+105
can be approximated by a `van der Laan' (1996) model for an expanding
plasmon. However, as discussed in FP98 such a model
does not well describe the observed flat spectrum, which seems instead
to be better modelled by a partially self-absorbed conical jet of the
type developed for AGN (e.g. Blandford \& K\"onigl 1979; Reynolds
1982). 

More recently Kaiser, Sunyaev \& Spruit (2000) have further applied a
internal shock model to the major 1994 radio outburst of the source
reported in MR94. In their model they require approximately the same
amount of energy to be associated with the events as calculated for a
plasmon model in MR94, but the {\em power} requirement is less as
they spread the energy input over a much longer period.  However, with
repeated oscillations as observed here, this cannot be the case, where
an entire accretion -- ejection cycle is repeated on the timescale of
$\sim 20$ min which we have used to calculate $P$ and $\dot{M}_{\rm
jet}$ in table 1. Therefore this model cannot be used to evade the
enormous amount of continuous power required to generate the observed
repeated ejection events (this is not an argument against their model,
but one against using it to evade the huge power requirements).
Importantly, unless 
(a) there is a bright mm--infrared contribution from the large ejections
which has not to date been observed, and (b) it is only the
spatially resolved ejections (F99; RM99) which have bulk relativistic
motion and a baryonic content, then {\em GRS 1915+105 injects more
energy and matter into the outflow during periods of repeated small
events than it does during the large ejections.}

Note also that Belloni, Migliari \& Fender (2000) have found that jet
power, calculated as above, appears to be anticorrelated with
accretion rate as inferred from X-ray spectral fits, for a small
sample of observations with quasi-simultaneous infrared and X-ray
coverage.

\section{Conclusions}

We have reported giant repeated oscillation events from the black hole
system GRS 1915+105 observed simultaneously at mm and infrared
wavelengths. Contemporaneous radio observations indicate that these
observations were near the end of a sequence of $\ga 10$ days of
oscillations with $\sim 20$-min quasi-periods (i.e. $\ga 700$ discrete
ejection events).
We have investigated in depth the energy and mass
flow associated with such events, seeking to minimize the very large
required power. However the magnetic field has an upper limit imposed
by the lack of observed radiation losses in the infrared band, so that
reducing the effective volume by means of a small filling factor
cannot significantly reduce the required power. Given the repeated
nature of the events, and their almost certain association to an
accretion cycle, we cannot spread the minimum required energy over a
longer timescale than the repetition quasi-period. As a result we find
that at least $3 \times 10^{38}$ erg s$^{-1}$ is required to be
channelled into the formation of the ejecta; a very significant
fraction of the accretion energy unless the black hole is very massive
indeed. The enormous energy budget associated with baryonic ejections
at bulk relativistic velocities may rule out this possibility,
although at present we do not know which is more unlikely: no baryons
or low velocity ejections.

\section*{Acknowledgements}

We would like to thank John Davies,
Graeme Watt, Fred Baas, Ian Robson, Iain Coulson, Andy Adamson,
Garret Cotter, Will Grainger, Mark Lacey, Susan Ridgeway, Tim Carroll
and Thor Wold for assistance in the realisation of these observations, 
and Christian Kaiser for stimulating discussions.
The JCMT is operated by The Joint Astronomy Centre on
behalf of the UK Particle Physics and Astronomy Research Council
(PPARC), the Netherlands Organisation for Scientific
Research and the National Research Council of Canada.  UKIRT is
operated by The Observatories on behalf of the PPARC.  We thank the
staff at MRAO for maintenance and operation of the RT,
which is supported by the PPARC.

\end{document}